\documentclass[aip,
pof,
amsmath,
amssymb,
preprint,
numerical]{revtex4-1}
\usepackage[T1]{fontenc}
\usepackage[latin9]{inputenc}
\usepackage{xcolor}
\usepackage{pdfcolmk}
\usepackage{units}
\usepackage{graphicx}
\usepackage{esint}
\PassOptionsToPackage{normalem}{ulem}
\usepackage{ulem}

\makeatletter

\usepackage{bpchem}
\usepackage{siunitx}
\usepackage{color}

\newcommand{\bi}{ \boldsymbol}
\def\vec#1{\boldsymbol{\mathit{#1}}}

\makeatother

\begin{document}

\title{The influence of current collectors on Tayler instability and
  electro-vortex flows in liquid metal batteries}

\author{N. Weber}
\affiliation{ 
  Helmholtz-Zentrum Dresden - Rossendorf, Bautzner Landstr. 400, 01328
  Dresden, Germany
}%
\author{V. Galindo}
\affiliation{ 
  Helmholtz-Zentrum Dresden - Rossendorf, Bautzner Landstr. 400, 01328
  Dresden, Germany
}%
\author{J. Priede}
\affiliation{ 
  Engineering and Computing, Coventry University, Coventry, CV1 5FB United Kingdom
}%
\author{F. Stefani}
\affiliation{ 
  Helmholtz-Zentrum Dresden - Rossendorf, Bautzner Landstr. 400, 01328
  Dresden, Germany
}%
\author{T. Weier}
\email{t.weier@hzdr.de}
\affiliation{ 
  Helmholtz-Zentrum Dresden - Rossendorf, Bautzner Landstr. 400, 01328
  Dresden, Germany
}%

\begin{abstract}
The Tayler instability is a kink-type flow instability which occurs
when the electrical current through a conducting fluid exceeds a certain
critical value. Originally studied in the astrophysical context, the
instability was recently shown to be also a limiting factor for the
upward scalability of liquid metal batteries. In this paper, we continue
our efforts to simulate this instability for liquid metals within
the framework of an integro-differential equation approach. The original
solver is enhanced by multi-domain support with Dirichlet-Neumann
partitioning for the static boundaries. Particular focus is laid on
the detailed influence of the axial electrical boundary conditions
on the characteristic features of the Tayler instability, and, secondly,
on the occurrence of electro-vortex flows and their relevance for
liquid metal batteries. 
\end{abstract}
% http://www.aip.org/publishing/pacs/pacs-reg40#47
% 47.11.Df Finite volume methods
% 47.20.-k 	Flow instabilities
% 52.30.Cv Magnetohydrodynamics

\maketitle

\section{Introduction}

It is well known that the interaction of an electrical current with
its own magnetic field in a conducting fluid can create equilibrium
states that are not always stable. For example, the electrical discharge
in a plasma column can be prone either to the axisymmetric ``sausage''
or the non-axisymmetric ``kink'' instability. At the conductivity
and viscosity values typical for plasmas, such instabilities are common,
and special care must be taken to avoid them. For the case of a $z$-pinch,
and similarly for Tokamak geometry, this is expressed by the Kruskal-Shafranov
condition stating that the kink instability can only be suppressed
by applying an additional longitudinal magnetic field \cite{Goedbloed2004}.Things
are different for liquid metals, in which the high values of resistivity
and viscosity play a stabilizing role. For the paradigmatic case of
a homogeneous current through a cylindrical liquid metal column, this
type of instability is governed by the ratio of magnetic to viscous
forces, expressed by the square of the Hartmann number 
\[
{\rm Ha}=B_{0}R(\sigma/\rho\nu)^{1/2},
\]
where $B_{0}=B_{\varphi}(R)$ is the azimuthal magnetic field at the
cylinder radius $R$; $\sigma$, $\rho$, and $\nu$ are the conductivity,
density and kinematic viscosity of the fluid. Beyond a critical value
of ${\rm Ha}\simeq20$, see R\"udiger et al.\cite{Ruediger2011}, a non-axisymmetric flow
pattern will appear in the fluid reaching a mean velocity value $V$
whose corresponding Reynolds number ${\rm Re}=RV/\nu$ scales as ${\rm Re}\sim{\rm Ha}^{2}$.

In contrast to the vast experience with the kink-type instability
in plasma physics, its occurrence in a liquid metal was experimentally
demonstrated only recently \cite{Seilmayer2012}. This instability,
which needs a certain critical current to overcome the stabilizing
effect of resistivity and viscosity, is referred to as the \textit{Tayler
instability} (TI) -- the term originally restricted to the destabilizing
effect of a magnetic field on a stably stratified fluid in cosmic
bodies. In this astrophysical context, the TI had been discussed in
connection with a non-linear stellar dynamo mechanism \cite{Spruit2002},
and the fast slow-down of white dwarfs \cite{Suijs2008}.

\begin{figure}[t]
\centering{}\includegraphics[width=6cm]{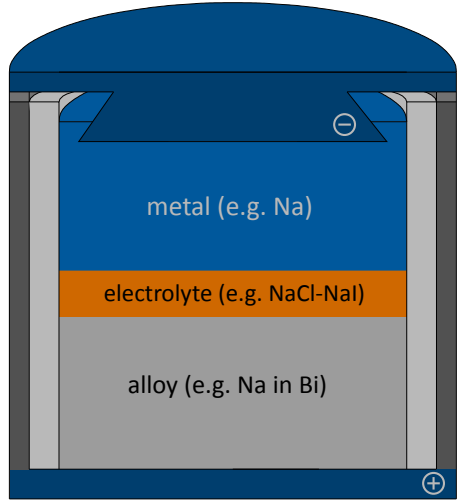}
\caption{\label{fig:battery}Sketch of a typical Liquid Metal Battery (LMB).}
\end{figure}

Apart from its astrophysical significance, the TI may also be relevant
for the applied problem connected with the revived interest in liquid
metal batteries (LMB), which are promising as a relatively cheap storage
for the highly fluctuating renewable energies \cite{Kim2013b}. Liquid
metal batteries consist basically of a self-assembling stratified
system of a heavy liquid metal or metalloid (e.g., Bi, Sb) at the
bottom, a molten salt mixture as electrolyte in the middle, and a
light alkaline or earth alkaline metal (e.g., Na, Mg) at the top (see
Fig.~\ref{fig:battery}). Choosing \BPChem{Na} and \BPChem{Bi}
as an example, \BPChem{Na} will loose one electron during the discharge
process, turning into \BPChem{Na\^{+}}. This ion diffuses through
the electrolyte into the lower \BPChem{Bi} layer where it is reduced
and alloys with \BPChem{Bi} to \BPChem{NaBi}.

To be economically competitive, such batteries should be built rather
large. With envisioned current densities of up to $\unit[10]{kA/m^{2}}$, a base area of $\approx\unit[1]{m^{2}}$ would immediately lead
to total currents of several kA, which poses a certain risk of the
TI. The TI-driven flow would then endanger the integrity of the (poorly
conducting) electrolyte layer \cite{Weber2014}, which should usually
be chosen as thin as possible in order to reduce energy losses due
to Joule heating. Despite the fact that it is well possible to suppress
the TI by either using a back-directed current in the centre \cite{Stefani2011},
or by invoking the Kruskal-Shafranov idea, i.e. by applying an additional
axial magnetic field \cite{Weber2014}, a more detailed knowledge
of the characteristics of the TI in liquid metal batteries is still
desirable.

Although strong fluid flows represent a danger for the LMB, a slight
flow in the cathode compartment may be advantageous. Several researchers
observed concentration polarisation \cite{Kim2013b,Heredy1967,Agruss1967,Foster1967b,Cairns1967}
caused by mass transfer limitations while (dis-)charging LMBs. These
may be due to, for example, a concentration of reaction products at
the electrolyte-alloy interface. Here, a careful mixing of the alloy
would be favourable \cite{Foster1967b,Cairns1967}, e.g. by exploiting
electro-vortex flow or thermal convection \cite{Kelley2014}.

In a recent paper \cite{Weber2013}, we made a first attempt to simulate
the TI for realistic material parameters of liquid metals. Rather
than applying the usual method of solving the coupled system of Navier-Stokes
equation for the velocity and induction equation for the magnetic
field, we utilized an integro-differential equation approach in which
the hydrodynamics is still simulated by the Navier-Stokes equation
while the electromagnetic part is solved by applying a Poisson equation
for the electric potential and the Biot-Savart law for the magnetic field.
With this method we were able to confirm the experimental results
of Seilmayer et al.\cite{Seilmayer2012}, and to predict the critical currents for
onset of the TI for varying aspect ratio.

In this first approach we assumed the initial current distribution
to be strictly vertical and the magnetic field to depend spatially
only on the radius. Also the axial electrical boundary conditions
at the current conductors were simplified by setting the electric
potential to a constant on the whole interface, which corresponds
to perfectly conducting solid current collectors. Using this approach
we determined only the current perturbation in the liquid but ignored
the currents closing through the well conducting collectors. Application
of Biot-Savart law to open current loops resulted in not entirely
correct distribution of the magnetic field perturbation. 

In the present paper, we aim at clarifying the previously neglected
effect of the current distribution in the current collectors on the
characteristics of the TI. First, we extend the numerical scheme of
Weber et al.\cite{Weber2013} by including iterative solution of the Laplace equation
for the electric potential in the solid current collectors. In a
second step, we focus on the initial current and magnetic field distribution.
We solve a Laplace equation for the electric potential in the whole
domain in order to obtain the initial current distribution. Compared
to the simplified approach of Weber et al.\cite{Weber2013}, we observe now also
radial currents, driving an additional electro-vortex flow.

As a main result we show that an appropriate design of current collectors
is very important in regard to the resulting electro vortex flow.
The effect of missing current collectors on the TI is often considerable,
but in certain cases it will be justified to restrict the attention
to the numerics in the liquid. The developed method is a first step
towards a detailed investigation of the coupled problem of TI, electro-vortex flow, thermal convection and interfacial instabilities at the
boundaries between the three layers of a liquid metal battery.

\section{Mathematical Model}

In this section, we will basically reiterate the derivation of the
integro-differential equation approach as it was developed by Weber et
al.\cite{Weber2013}
for the numerical simulation of the TI, but will enhance it with a
description of the treatment of the electric potential in the solid
current collectors.

The fluid dynamics in a liquid metal is governed by the Navier-Stokes
equation (NSE) for the velocity $\vec{v}$, 
\begin{equation}
\partial_{t}\vec{v}+\left(\vec{v}\cdot\vec{\nabla}\right)\vec{v}=-\frac{1}{\rho}\vec{\nabla}p+\nu\Delta\vec{v}+\frac{1}{\rho}\vec{f},\label{eqn:navierstokes}
\end{equation}
where $p$ denotes the pressure, $\rho$ the density, $\nu$ the kinematic
viscosity, and ${\bi f}$ the body force density. Further, the continuity
equation $\vec{\nabla}\cdot\vec{v}=0$ is needed to ensure the incompressibility
of the fluid.

Consider now an electrical current density $\bi J$ in the fluid,
together with the magnetic field ${\bi B}$ related to it via \mbox{$\mu_{0}\bi J=\nabla\times{\bi B}$},
with $\mu_{0}$ being the permeability of the free space (magnetic
materials are not treated in this paper). The cross product of $\bi J$
and ${\bi B}$ gives the Lorentz force, which is the only body force
to be considered in the following: 
\begin{equation}
\bi f=\bi f_{\mathrm{L}}=\bi J\times\bi B.
\end{equation}
The total magnetic field ${\bi B}$ consists of two parts: the static
field $\bi B_{\mathrm{0}}$, generated by the externally applied current
$I$ (or the corresponding current density $\bi J_{\mathrm{0}}$),
and the induced magnetic field ${\bi b}$, generated by the flow induced
current ${\bi j}$. It is essential to note that in our problem ${\bi b}$
must not be neglected in the expression for the Lorentz force since
its product with the applied current ${\bi J_{\mathrm{0}}}$ is of
the same order as the product of $\bi B_{\mathrm{0}}$ with the induced
current ${\bi j}$.

A rigorous way to compute the magnetic field evolution is by solving
the induction equation 
\begin{equation}
\partial_{t}\vec{B}+(\vec{v}\cdot\vec{\nabla})\vec{B}=\left(\vec{B}\cdot\vec{\nabla}\right)\vec{v}+\frac{{\displaystyle 1}}{{\displaystyle \mu_{0}\sigma}}\Delta\vec{B}.
\end{equation}
The solution of this equation requires suitable boundary conditions
for the magnetic field, which can be implemented either by solving
a Laplace equation in the exterior of the fluid \cite{Kluwer1999,Guermond2007},
or by equivalent boundary element methods \cite{Stefani2000,Iskakov2004,Giesecke2008}.

In the following, we adopt a method proposed by Meir et al. \cite{Meir2004}
for the quasi-stationary case, i.e. we compute the magnetic field
using the Biot-Savart law 
\begin{equation}
\vec{b}(\vec{r})=\frac{\mu_{0}}{4\pi}\int dV'\,\frac{\vec{j}(\vec{r}')\times(\vec{r}-\vec{r}')}{\left|\vec{r}-\vec{r}'\right|^{3}},\label{eqn:biotsavart}
\end{equation}
which is the inversion of Ampère's law $\vec{\nabla}\times\vec{b}=\mu_{0}\vec{j}$.
Therefore, the problem is shifted from the determination of the magnetic
field $\bi b$ to the determination of the current density $\bi j$.

\begin{figure}[t]
\centering{}\includegraphics[width=8cm]{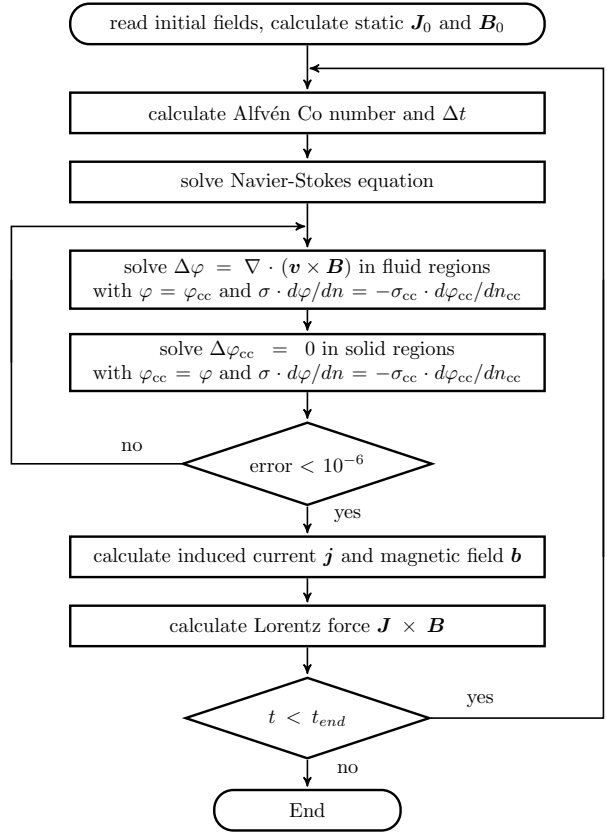}
\caption{\label{fig:workflow}Flow chart of the simulation model.}
\end{figure}

The flow chart of the numerical algorithm (Fig.~\ref{fig:workflow})
is similar to that worked out by Weber et al.\cite{Weber2013}, with the exception
that we need an additional loop to match the solution for the electric
potential in the current collectors to that in the liquid metal. Assuming
the current $I$ through our cylindrical vessel as given (e.g., the
battery charging current), we compute the corresponding current density
$\bi J_{\mathrm{0}}$, and the associated static magnetic field $\bi B_{\mathrm{0}}$.
In a first step, we assume the current density to be constant in the
whole domain and the magnetic field as resulting from an infinite
long cylinder. This allows us to study the influence of the current
collectors on the induced currents only. Thereafter, we also consider
the feeding lines of the battery by imposing a constant electric potential
on the interface between current collector and feeding line and solving
a Laplace equation in the solid and liquid domains. The static magnetic
field $\bi B_{\mathrm{0}}$ is computed by using the Biot-Savart law.
The field of the (infinitely long) feeding lines is added. For that
case, the modified current distribution has radial components, as
well, and the magnetic field slightly declines close to the current
collectors. Previously, the static Lorentz Force $\bi J_{0}\times\bi B_{0}$
had only radial components -- now axial components appear, driving
electro-vortex flows in the vicinity of the current collectors.

In the main loop of our numerical scheme, the Navier-Stokes equation
(\ref{eqn:navierstokes}) is solved first, followed by a velocity
corrector step to ensure continuity ($\vec{\nabla}\cdot\vec{v}=0$).
Then we have to find the electric current density, both in the liquid
metal where $\vec{v}\ne0$, and in the solid current collectors where
$\vec{v}=0$. Assuming the magnetic Reynolds number ${\rm Rm}=\mu_{0}\sigma Rv$
on the basis of the TI-triggered velocity scale $v$ to be small,
we can invoke the quasistatic approximation \cite{Davidson2001}.
This means that we express the electric field by the gradient of an
electrostatic potential, ${\bi E}=-\nabla\Phi$. Applying the divergence
operator to Ohm's law in moving conductors, ${\bi j}=\sigma(-\nabla\Phi+{\bi v}\times{\bi B})$,
and demanding charge conservation, $\nabla\cdot{\bi j}=0$, we arrive
at the Poisson equation 
\begin{equation}
\Delta\varphi=\vec{\nabla}\cdot\left({\bi v}\times{\bi B}\right)
\end{equation}
for the perturbed electric potential $\varphi=\Phi-\Phi_{0}$, where
$\Phi_{0}$ is the potential according to the externally applied current.
Since no current can flow through the insulating rim of the cylindrical
vessel, we assume here $\bi n\cdot\vec{\nabla}\varphi=0$, where $\bi n$
is the surface normal vector.

The new point compared to Weber et al.\cite{Weber2013} is now that we have to
consider two regions with different conductivities. In the liquid
metal, with conductivity $\sigma$, we have to solve the Poisson equation
with $\vec{\nabla}\cdot({\bi v}\times{\bi B})$ as the source term.
In the solid current collector region, with conductivity $\sigma_{cc}$,
this source term is absent, and we have to solve the Laplace equation
$\Delta\varphi_{cc}=0$. What is still needed are the continuity conditions
for the potential and the normal current component at the interface
between the two regions. The fastest convergence for this Dirichlet-Neumann
partitioning is obtained by combining the two boundary condition to
a single Dirichlet condition for the interface potential 
\begin{equation}
\varphi_{i}=\frac{\varphi_{cc}\sigma_{cc}/\delta_{cc}+\varphi\sigma/\delta}{\sigma_{cc}/\delta_{cc}+\sigma/\delta}
\end{equation}
with $\varphi_{cc}$ and $\varphi$ denoting the potential in the
cell centres next to the boundary and $\delta_{cc}$ and $\delta$ the distance
between cell centre and boundary. The subscript $cc$ stands here
for the solid current collector, whereas the variables without subscripts
refer to the liquid metal region. The convergence criterion is $10^{-6}$,
where the better conducting region is taken for determining the relative
error.

Having derived the potential in the fluid by solving the Poisson equation,
we easily recover the current density induced by the fluid motion
as 
\begin{equation}
{\bi j}=\sigma\left(-\nabla\varphi+{\bi v}\times{\bi B}\right)
\end{equation}
and then the induced magnetic field by using the Biot-Savart law (\ref{eqn:biotsavart}).
As an improvement to Weber et al.\cite{Weber2013}, we include now also the current
distribution in the current collectors. The total current density
is 
\begin{equation}
{\bi J}=\bi J_{0}+\bi j.
\end{equation}
In a last step, the Lorentz force, to be implemented into the Navier-Stokes
equation, is calculated according to 
\begin{equation}
\bi f_{\mathrm{L}}=\left(\sigma\left(-\vec{\nabla}\varphi+{\bi v}\times\left(\bi B_{\mathrm{0}}+{\bi b}\right)\right)+\bi J_{0}\right)\times\left(\bi B_{\mathrm{0}}+{\bi b}\right).
\end{equation}
It should be noticed again that the (small) induced magnetic field
$\bi b$ cannot be omitted here, since its product with the (large)
imposed current $\bi J_{0}$ is of the same order as the product of
the (small) induced current $\bi j$ with the (large) magnetic field
$\bi B_{0}$.

\section{Exploring the role of current collectors}

In this section we will study in detail the effect of the current
distribution in the solid current collectors on the TI. We chose a
liquid metal column of constant height $H=\unit[1.2]{m}$, diameter
$D=2R=\unit[1]{m}$, aspect ratio $H/D=1.2$ and electric conductivity
$\sigma$. We apply an electrical current corresponding to ${\rm Ha}=250$
and vary the height $H_{cc}$ as well as the conductivity $\sigma_{cc}$
of the solid current collectors. Remember that the latter ones were
not included by Weber et al. \cite{Weber2013} where the simplified condition $\varphi=\mbox{\emph{const}}$
was chosen as interface condition at top and bottom of the liquid
metal column, and the currents in the current collectors were also
ignored in the Biot-Savart law. From physical arguments, we anticipate
that in long liquid metal columns the effect of the detailed current
distribution at the ends of the current collectors should be negligible,
but that it might still be significant in shallow layers.

\begin{figure}[t]
\centering{}\includegraphics[width=\textwidth]{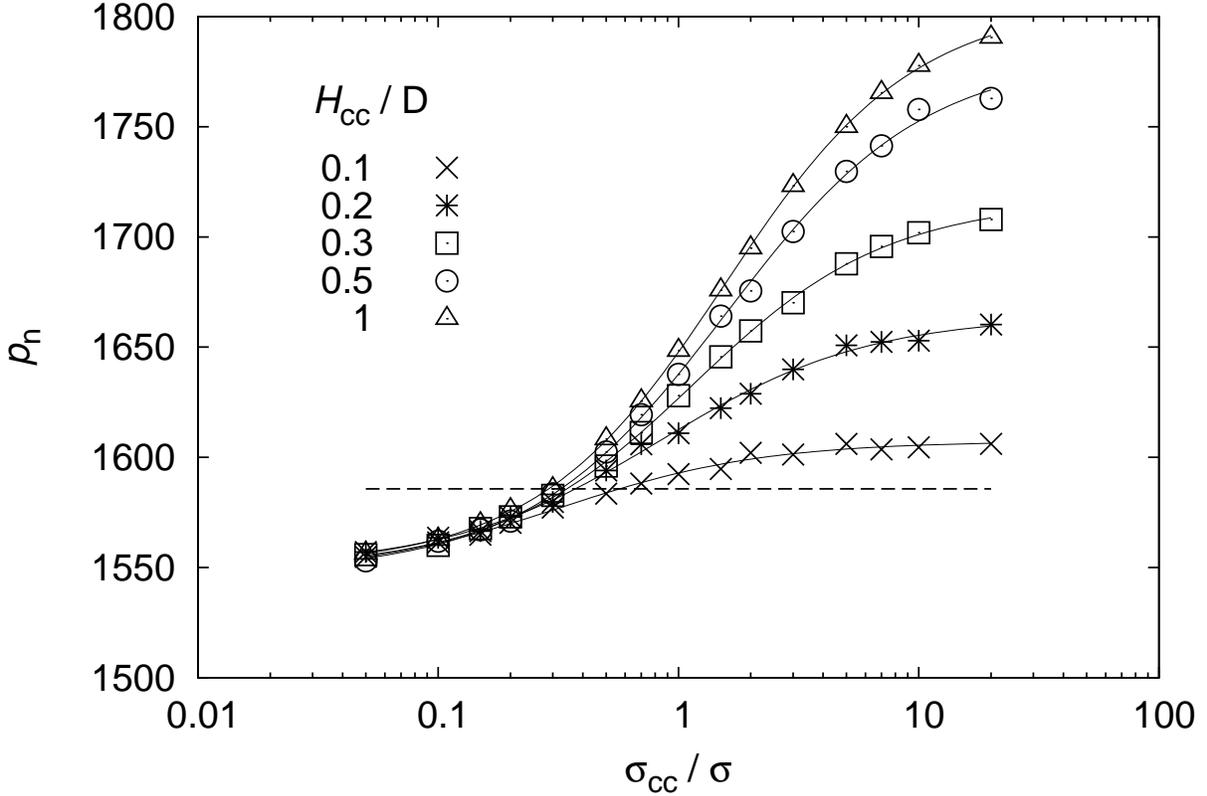}
\caption{\label{fig:gr}The growth rate of the TI against the conductivity
ratio of current collectors of various height for a cylindrical liquid
metal column with aspect ratio $H/D=1.2$, diameter $D=\unit[1]{m}$
and ${\rm Ha}=250$. The physical growth rate is normalised by the
inverse of the viscous time scale $R^{2}/\nu$. The dashed line is
the growth rate for a simulation with the equipotential boundary conditions
and ignored current perturbation in the collectors. }
\end{figure}

We expect the current collectors to influence the magnitude of the
fluid flow as well as the growth rate of the TI. Figure \ref{fig:gr}
shows the latter for current collectors of various height and an electric
conductivity of $0.05$ up to $20$ times of the liquid metal. The
height of the current collector varies from $10$ to $\unit[100]{\%}$
of the diameter.
 
\begin{figure}[t]
\centering{}\includegraphics[width=\textwidth]{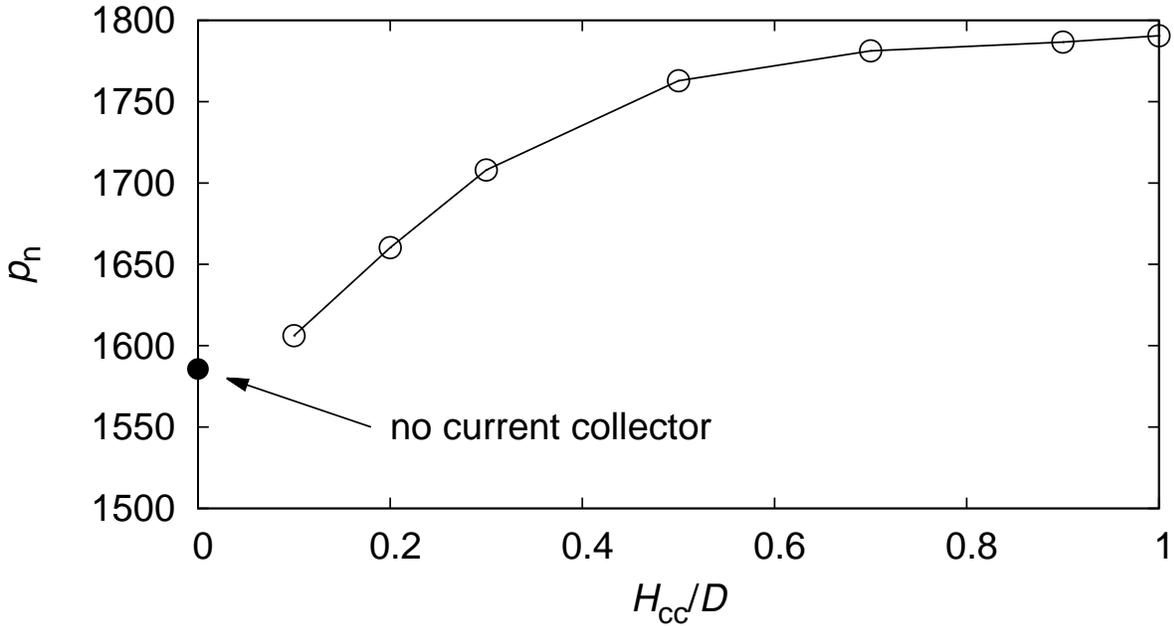}
\caption{\label{fig:grmax}The growth rate of the TI normalised by the inverse
of the viscous time scale $R^{2}/\nu$ against the height of current
collector for a cylindrical liquid metal column with aspect ratio
$H/D=1.2,$ diameter $D=\unit[1]{m},$ the applied current corresponding
to ${\rm Ha}=250,$ and the conductivity ratio of solid and liquid
regions $\sigma_{cc}/\sigma=20$. }
\end{figure}

The resulting curves are fitted well by an arctangent function with
the constants depending on conductivity and aspect ratio of the current
collector. Choosing a poorly conducting current collector, the growth
rate approaches an asymptotic minimum (Fig.~\ref{fig:gr}, left).
This is the same value for all current collectors, regardless of their
aspect ratio. We will show later, that current collectors of low conductivity
force the current loops to close already within the liquid metal.
Very little current perturbation spreading into the poorly conducting
current collector makes the instability insensitive to the height
of the collector.

Turning to the current collectors of a high relative conductivity
compared to the liquid metal (Fig.~\ref{fig:gr}, right) we observe
that the growth rate tends to an asymptotic maximum as the height
of the current collector becomes comparable to the diameter of the
cylinder. Obviously, this is a typical vertical distance over which
the currents close in the solid current conductors. The magnetic field
of these currents slightly increases the growth rate of the TI. In
Fig.~\ref{fig:grmax} we show again the growth rate of the TI, but
for a fixed conductivity ratio of $\sigma_{cc}/\sigma=20$, i.e. for
very well conducting current collectors. With increasing the height,
the growth rate increases, too, approaching an asymptotic maximum.
A well conducting current collector which is as high as wide allows
almost all currents to close. Further increase of the collector height
has almost no effect on the TI.

Another important aspect is the comparison to the growth rate of the
(old) simulation without current collectors (Fig.~\ref{fig:gr},
dotted line and Fig.~\ref{fig:grmax}, black dot). As we will show
later, omitting the current collectors has no grave effect on
the current perturbation in the liquid metal. The TI is affected by
the current perturbation penetrating into the well conducting current
collectors which was ignored in the previous simulation. The effect
of this omission is illustrated by the asymptotic value for an infinitely
flat current collector (aspect ratio $H_{cc}/D=0$) shown in Fig.~\ref{fig:grmax}.

\begin{figure}[t]
\centering{}\includegraphics[width=\textwidth]{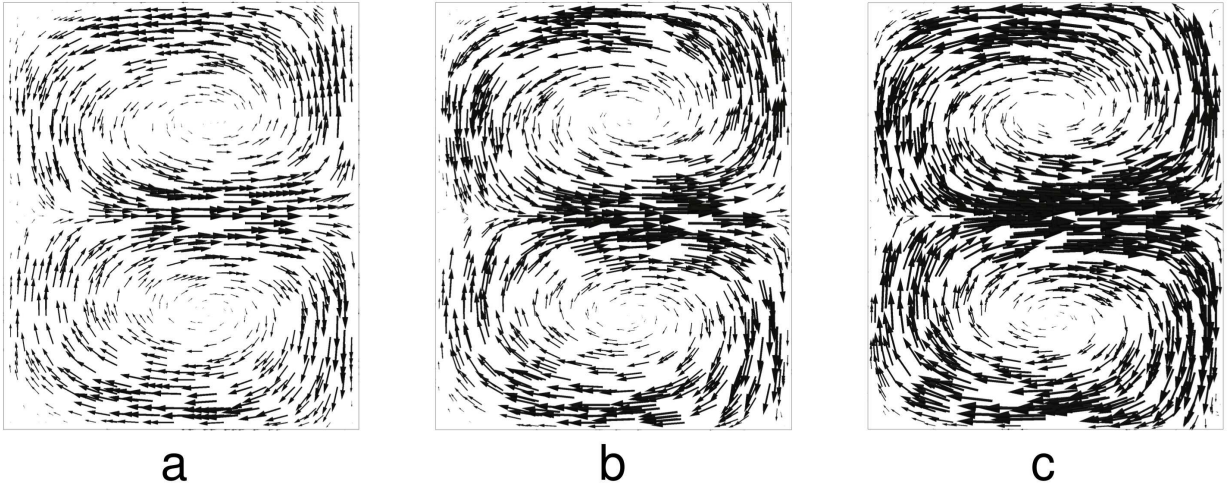}
\caption{\label{fig:u}Velocity field of a saturated TI in a cylinder with
$D=\unit[1]{m},$ $H=\unit[1.2]{m}$ and ${\rm Ha=250}$ without current
collector (a) and current collectors with aspect ratio $H_{cc}/D$
of 0.1 (b) and 0.5 (c) and constant conductivity ratio $\sigma_{cc}/\sigma=5$.}
\end{figure}

\begin{figure}[t]
\centering{}\includegraphics[width=\textwidth]{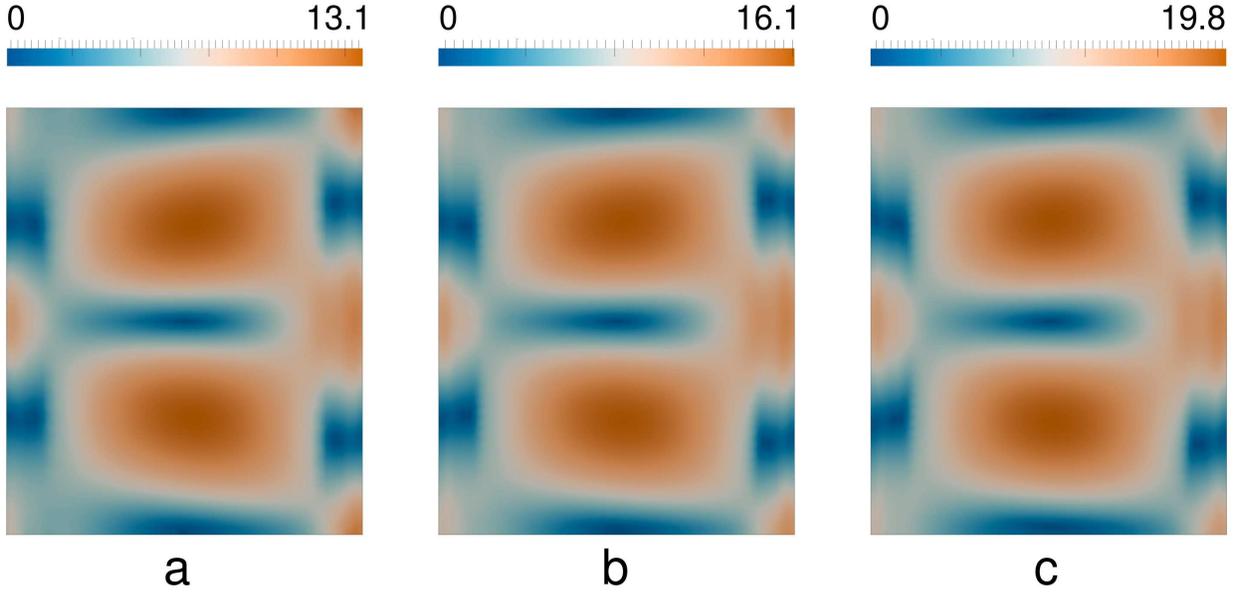}
\caption{\label{fig:j}Magnitude of the induced current density in A/m$^{2}$
of the saturated TI in a cylinder with $D=\unit[1]{m},$ $H=\unit[1.2]{m}$
and ${\rm Ha=250}$ without current collector (a) or a current collector
of aspect ratio $H_{cc}/D$ of 0.1 (b) and 0.5 (c). The conductivity
ratio between solid and liquid region is $\sigma_{cc}/\sigma=5$.}
\end{figure}

\begin{figure}[t]
\centering{}
\includegraphics[width=\textwidth]{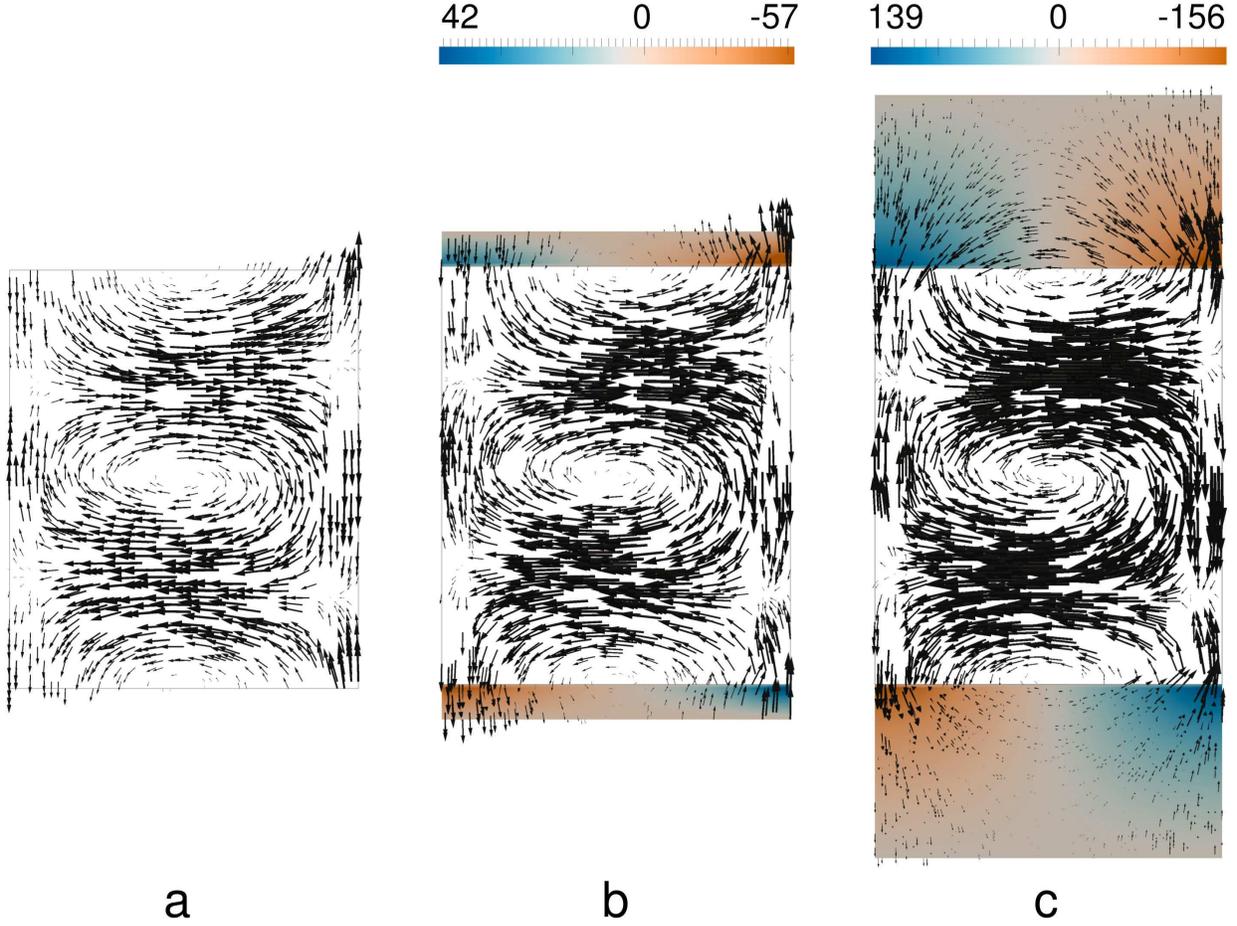}
\caption{\label{fig:potAndJ}Electric potential perturbation (nV) in the current collectors and
the induced current lines in the saturated TI. The geometry consists
of a cylindrical liquid metal column with $D=\unit[1]{m}$, $H=\unit[1.2]{m}$
without (a) or with a current collector of aspect ratio $H_{cc}/D=0.1$
(b) and 0.5 (c). For the last two cases the current collectors are
five times better conducting than the fluid ($\sigma_{cc}/\sigma=5$.)}
\end{figure}

We will now discuss some exemplary cases in order to illustrate our
preceding argumentation. First of all, we will explore the effect
of the aspect ratio of the current collectors for the conductivity
ratio $\sigma_{cc}/\sigma=5.$ Three cases are considered: without
current collector (a), with a flat current collector (b) and a high
one (c) (see Figs.~\ref{fig:u},\ref{fig:j} and \ref{fig:potAndJ}).
In the following, we will focus on the saturated TI and ignore the
considerably different flow fields which may sometimes be observed
during the exponential growth phase of the TI due to spontaneous symmetry
breaking \cite{Bonanno2012}.

Figure \ref{fig:u} illustrates the velocity field for the three paradigmatic
cases. We observe clearly a globally increasing velocity when adding
a current collector (b) and rising its height (c). Interestingly,
the form of the convection cells does not change. Comparing Figs.~\ref{fig:j}a and c, an increase of the induced current of approximately
$50\%$ can be observed when adding a current collector of aspect
ratio $H_{cc}/D=0.5$. Again, there is no obvious qualitative difference
between the simulation with (b,c) and without (a) current collector.
In Fig.~\ref{fig:potAndJ} we show now the current lines in the
liquid metal column and the current collectors as well, using the
same geometry as before. Omitting the current collectors (Fig.~\ref{fig:potAndJ}a), we can clearly see that the current lines do
not close, while they do partially so for flat current collectors
(Fig.~\ref{fig:potAndJ}b) and even better for high ones (Fig.~\ref{fig:potAndJ}c). Note also that the disturbed potential at the
interfaces is not constant as assumed previously \cite{Weber2013}
and its nonuniformity increases considerably with increasing current
collectors aspect ratio (Fig.~\ref{fig:potAndJ}b and c).

\begin{figure}[t]
\centering{}
\includegraphics[width=\textwidth]{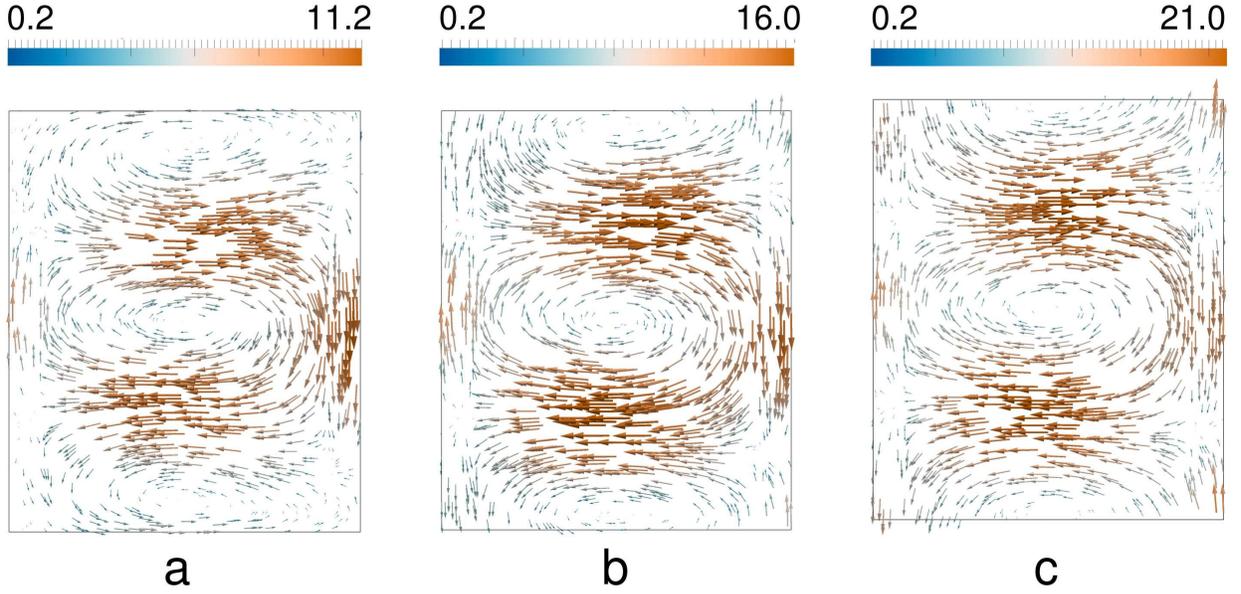}
\caption{\label{fig:onlySigma}Induced current density in A/m$^{2}$ of the
saturated TI in a cylindrical liquid metal column ($D=1$\,m, $H=1.2$\,m)
with a current collector of constant height $H_{cc}=0.5$\,m and
an applied current corresponding to ${\rm Ha=250}$. The conductivity
ratio of solid to liquid region $\sigma_{cc}/\sigma$ is 1/20 (a),
1 (b) and 20 (c). The colour indicates the modulus of the current
density.}
\end{figure}

Having explored the role of the current collectors height, we will
focus now on its electric conductivity. We compare three current collectors
of fixed height $H_{cc}/D=0.5$ (see Fig.~\ref{fig:onlySigma}).
First, we consider a very poorly conducting solid current collector
with a conductivity of 1/20 of the liquid metal (Fig.~\ref{fig:onlySigma}a).
It is remarkable, that almost all currents close within the liquid
metal, with a concentration at the top and bottom of the cylinder.
When the conductivity of the current collector is increased to the
same value as of the liquid metal, there are much less currents closing
near the interface (Fig.~\ref{fig:onlySigma}b). A further elevation
of conductivity amplifies this effect (Fig.~\ref{fig:onlySigma}c).
At the same time, the maximum value of induced currents almost doubles.
While poorly conducting current collectors make the induced current
lines close within the liquid metal (Fig.~\ref{fig:onlySigma}a),
the equipotential boundary conditions, which effectively model perfectly
conducting collectors, result in the induced current lines running
in and out of the liquid metal through the top and bottom interfaces
(see Fig.~\ref{fig:potAndJ}a).

\begin{figure}[t]
\centering{}
\includegraphics[width=\textwidth]{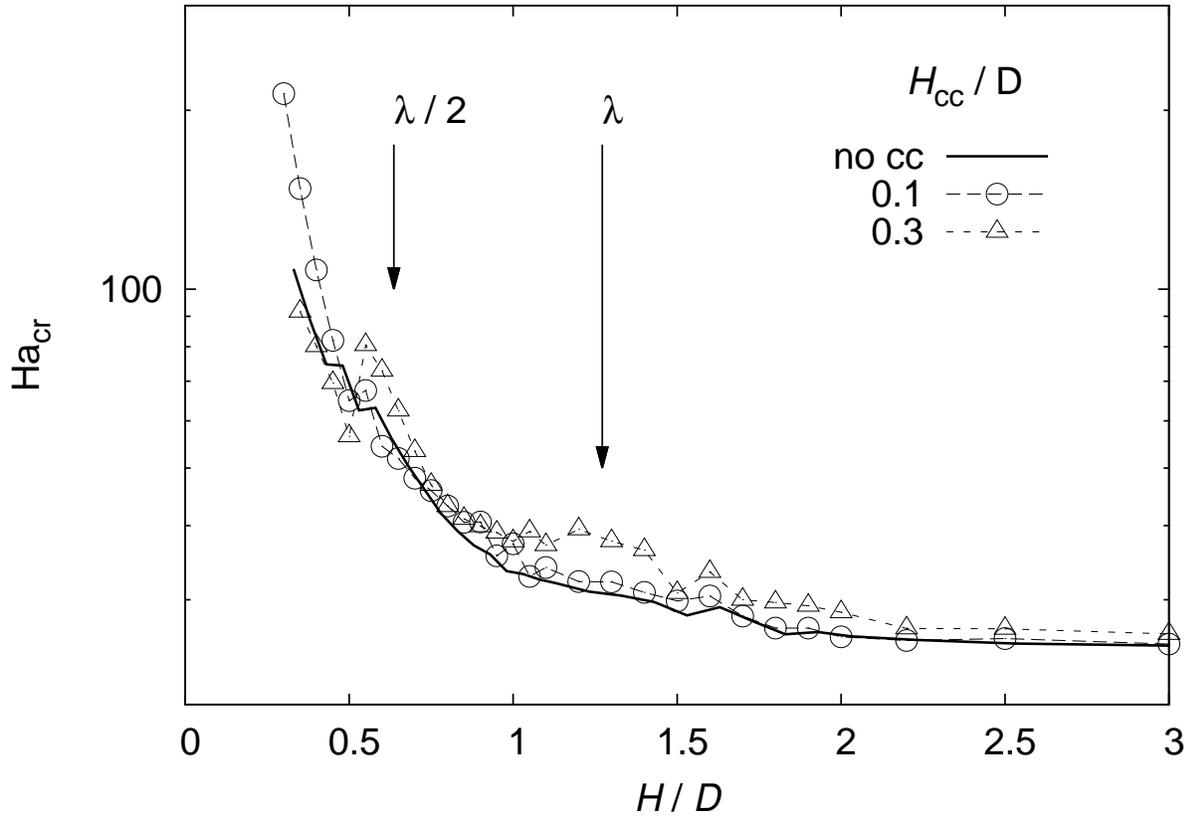}
\caption{\label{fig:ha}Critical Hartmann number for onset of the TI depending
on the aspect ratio of the liquid metal column. The current collectors
have twice the conductivity of the fluid ($\sigma_{cc}/\sigma=2$),
the diameter is $D=1$\,m.}
\end{figure}

Knowing the influence of current collectors on liquid metal columns
of aspect ratio \mbox{$H/D = 1.2$}, one may expect the effect to be much
stronger on very flat cells. On the other hand, the current collectors
may be neglected for tall liquid metal columns, as the influence should
be limited to the top and bottom regions with length comparable to
the diameter of the cylinder. In order to confirm this, we plot the
critical Hartmann number for onset of the TI for a liquid metal column
of different height (Fig.~\ref{fig:ha}). The diameter of the column
($D=1$\,m) and the current collectors aspect ratio $H_{cc}/D=0.1;0.3$
is constant. At this point it should be noted that the number of convection
cells depends markedly on (very) small changes of the column height,
see also Weber et al.\cite{Weber2013}. The current collector will surely influence
the number of vortices, as well, and may thus induce irregularities,
like plateaus, in the plot. 

Comparing the new simulation for a cylindrical liquid metal column
(Fig.~\ref{fig:ha}, black line) to previous simulation for cubic
geometries \cite{Weber2013}, there are no plateaus of the critical
current any more. We assume these plateaus to be caused by the interaction
of the cubic geometry and the transition between different numbers
of convection cells. Returning to Fig.~\ref{fig:ha} we observe
an almost identical critical current for onset of the TI for very
high liquid metal columns, regardless of the current collectors. Interestingly,
very flat current collectors ($\circ$) slightly increase the critical
current, when flattening the liquid metal column. Although this effect
is not very distinct, it is not what we expect (see also Fig.~\ref{fig:gr}).
Higher current collectors ($\vartriangle$) show a very nonuniform
variation: for flat liquid metal columns they decrease the critical
current while increasing it clearly for an aspect ratio $H/D$ between
0.5 and 1.5. We suspect this behaviour to be caused by the transition
between different numbers of TI vortices in relation to the current
collectors. However, a detailed explanation has to be postponed.

\section{Exploring the role of the feeding lines}

\begin{figure}[t]
\centering{}
\includegraphics[width=\textwidth]{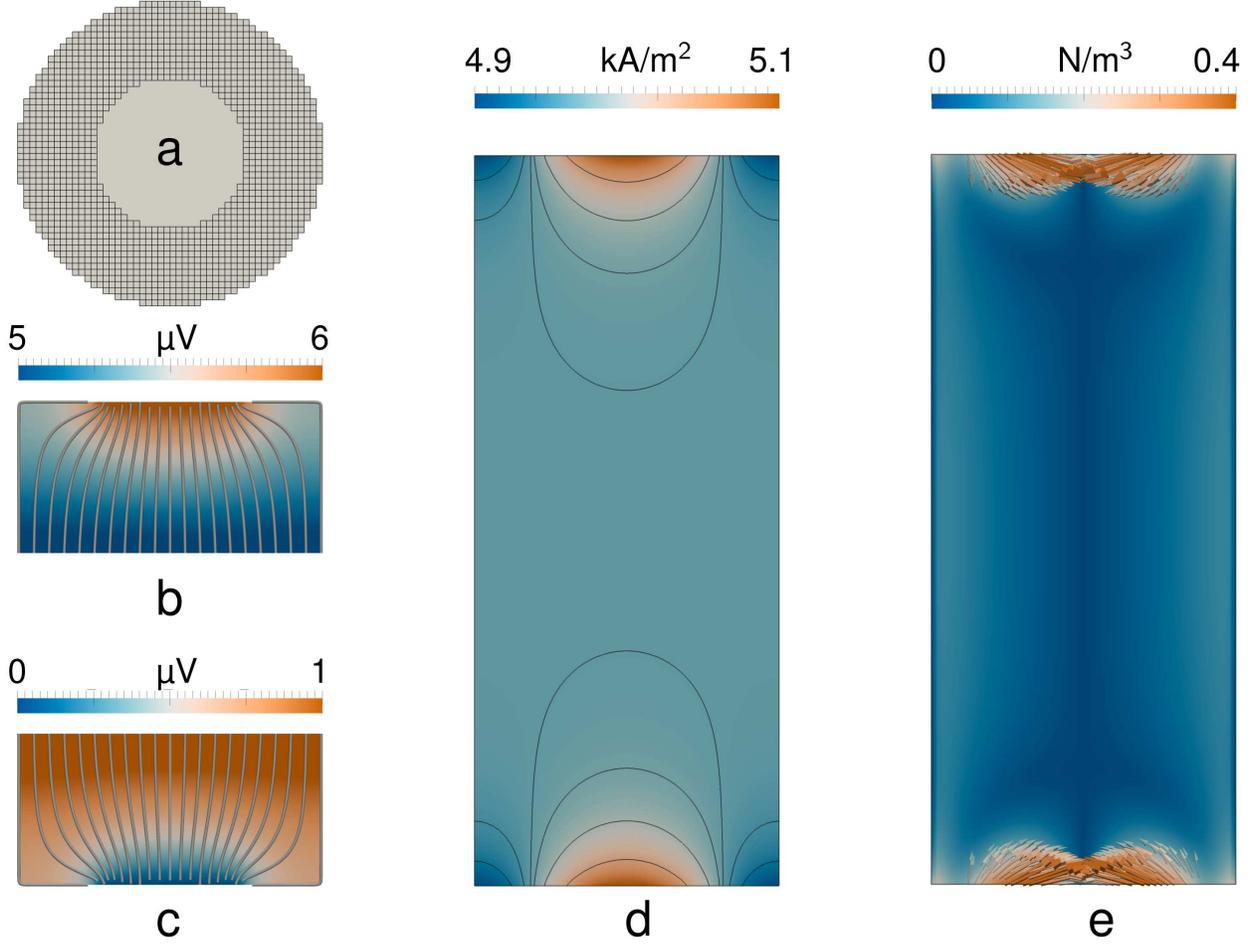}
\caption{\label{fig:fl:j}The top view of of the current collector with the
boundary condition $\varphi=0$ being applied on a circular area with
the diameter $d=D/2=0.5$\,m at the height $H_{cc}=0.5$\,m (a); the current pathes and the potential in the upper (b) and lower
(c) current collector; the current distribution in the liquid metal
(d) and the ``disturbed'' Lorentz force (e) calculated by subtracting
the Lorentz force for the infinite cylinder in order to visualise
the origin of the electro-vortex flow. Here the current collectors
have the same conductivity as the liquid metal column (of height $H=2.4$\,m)
and the applied current corresponds to ${\rm Ha=100}$.}
\end{figure}

In the preceding section we considered the induced current closing
in the current collector and investigated its influence on the growth
rate of the TI. Now, we add the feeding line above of the upper, and
below of the lower current collector. The diameter of the wires is
assumed to be half of that of the current collectors (see the grey
surface in Fig.~\ref{fig:fl:j}a). In the following we will consider
how electrically driven vortex flows arise and how to simulate them.
After describing interaction of these flows with the TI we will shortly
describe how they scale with the Hartmann number, the current collectors
aspect ratio and its electric conductivity.

In the previous section, the initial potential $P_{0}$ was determined
by setting $P_{0}=\textit{const}$ on the entire surface of the current
collector. Now, we use this boundary condition only on the interface
between the feeding line and the current collector (see the grey area
in Fig.~\ref{fig:fl:j}a). Thus, the static potential $P_{0}$ is
not longer radially uniform (see Figs. \ref{fig:fl:j}b and c).
This affects the base current density $\bi J_{0},$ which in our previous
model was purely vertical. Now, the current is purely vertical
and uniform only in the feeding line sufficiently far away from the
current collectors. As the current enters the current collector, it
spreads radially over the whole diameter (see the typical current structure
in Fig.~\ref{fig:fl:j}b for the upper and Fig.~\ref{fig:fl:j}c for
the lower current collector). Switching to the liquid metal column
in Fig.~\ref{fig:fl:j}d, we observe that the current is not completely
homogeneous. In Fig.~\ref{fig:fl:j}e we show the deviation of the
Lorentz force from its uniform distribution in an infinite liquid
metal column. The resulting force is no longer irrotational, as in the case of axially uniform
base current considered before, and thus it cannot be balanced by the
pressure gradient. This disbalance drives the so-called electro-vortex
flow \cite{Bojarevics1989}.

\begin{figure}[t]
\centering{}
\includegraphics[width=\textwidth]{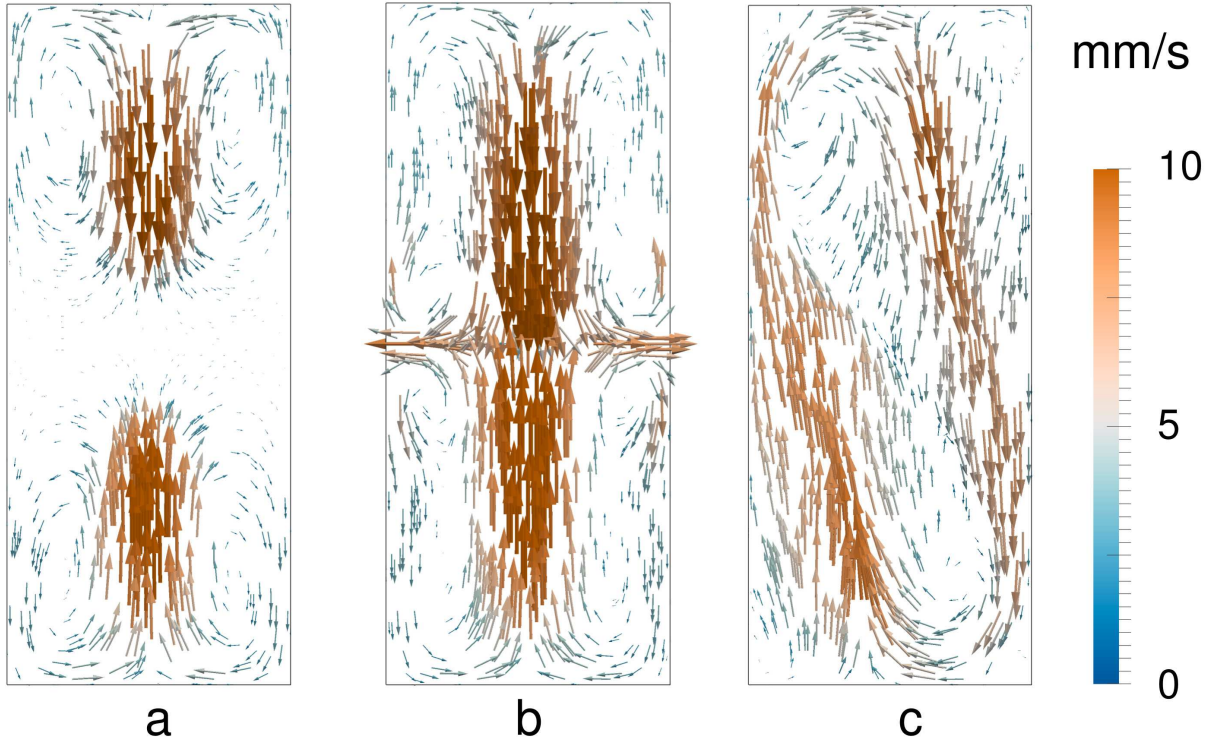}
\caption{\label{fig:fl:u}Temporal development of the electro-vortex flow.
The geometry consists of a cylindrical liquid metal column ($D=1$\,m,
$H=2.4$\,m) with two current collectors ($H_{cc}=0.5$\,m, $\sigma_{cc}/\sigma=5$)
and an applied current corresponding to ${\rm Ha=100}$. The original
$m=0$ mode (a: after 200\,s) becomes unstable (b: after 400\,s)
and turns into a $m=1$ mode (c: after 2300\,s).}
\end{figure}

In contrast to the TI, there is no threshold current for the onset
of electro-vortex flows. Although electro-vortex flow starts as soon
as the current is applied, some time is required for the flow to develop
through the whole volume (Fig.~\ref{fig:fl:u}a-b). As one would
deduce from the driving Lorentz forces (Fig.~\ref{fig:fl:j}e),
the initial flow structure is always axisymmetric, i.e., $m=0$ mode.
However, after some time, the electro-vortex flow often turns into
a wind (see Fig.~\ref{fig:fl:u}c). The time of transition 
depends not only on the applied current, but also on the aspect ratio
and conductivity of current collectors and liquid metal column. 

\begin{figure}[t]
\centering{}
\includegraphics[width=\textwidth]{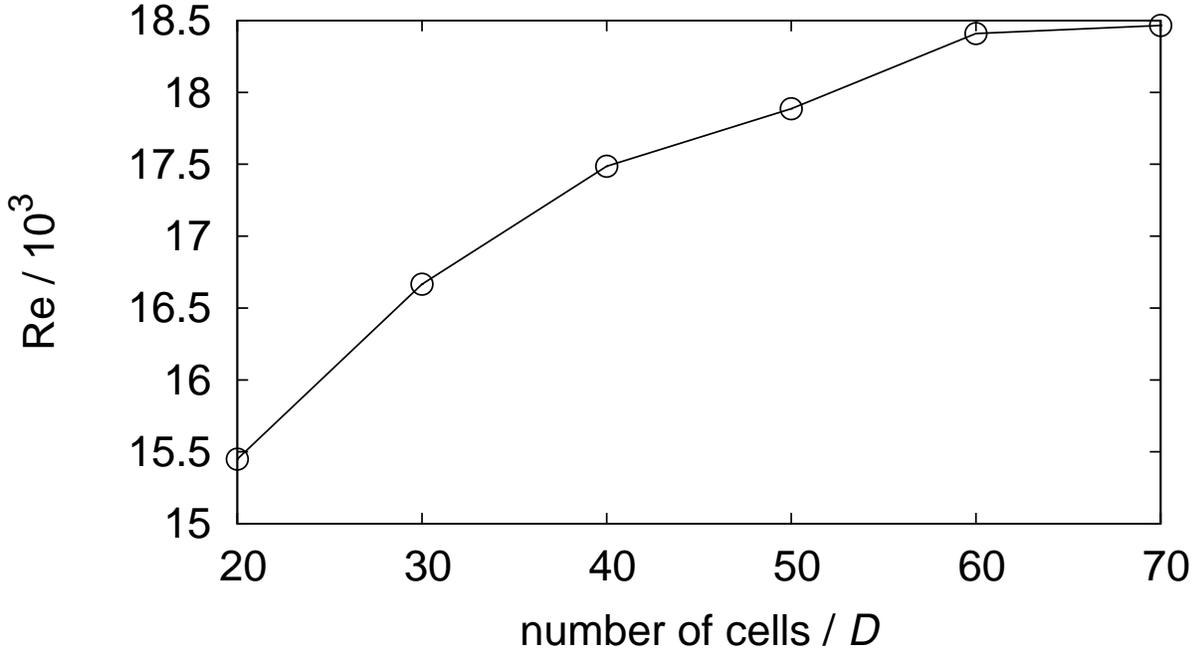}
\caption{\label{fig:gridstudy}Reynolds number associated with the electro-vortex
flow in a cylindrical liquid metal column ($D=1$\,m, $H=2.4$\,m)
with a current collector ($H_{cc}=0.5$\,m, $\sigma_{cc}/\sigma=5$)
and an applied current corresponding to ${\rm Ha=100}$ for varying
grid resolution.}
\end{figure}

Similar to thermal convection, electro-vortex flows are often strongly
time-dependent. Such simulation requires a much finer grid resolution
than used for the TI by Weber et al.\cite{Weber2013}. If we want to study the
interaction of TI and electro-vortex flow, the latter should be rather
week to allow the TI to develop. Bearing these two things in mind,
we choose the following simulation geometry: a relatively high liquid
metal column ($D=1$\,m, $H=2.4$\,m) of aspect ratio 2.4 together
with also high current collectors ($H_{cc}>0.5$\,m). This configuration
leaves enough ``space'' for the TI to develop, especially in the
middle of the cylinder and ensures a stationary flow. In Fig.~\ref{fig:gridstudy}
we show the dependence of the Reynolds number on the grid resolution
for this set-up. For our simulation we choose 25 purely cubic cells
over the radius. The quadrature of the circle ensures a good azimuthal
balance of the Lorentz forces (axial symmetry) (see also Fig.~\ref{fig:fl:j}a).
Flattening the current collectors below an aspect ratio of 0.5 increases
the Lorentz forces further and would definitely lead to an instationary
electro-vortex flow.

\begin{figure}[t]
\centering{}
\includegraphics[width=\textwidth]{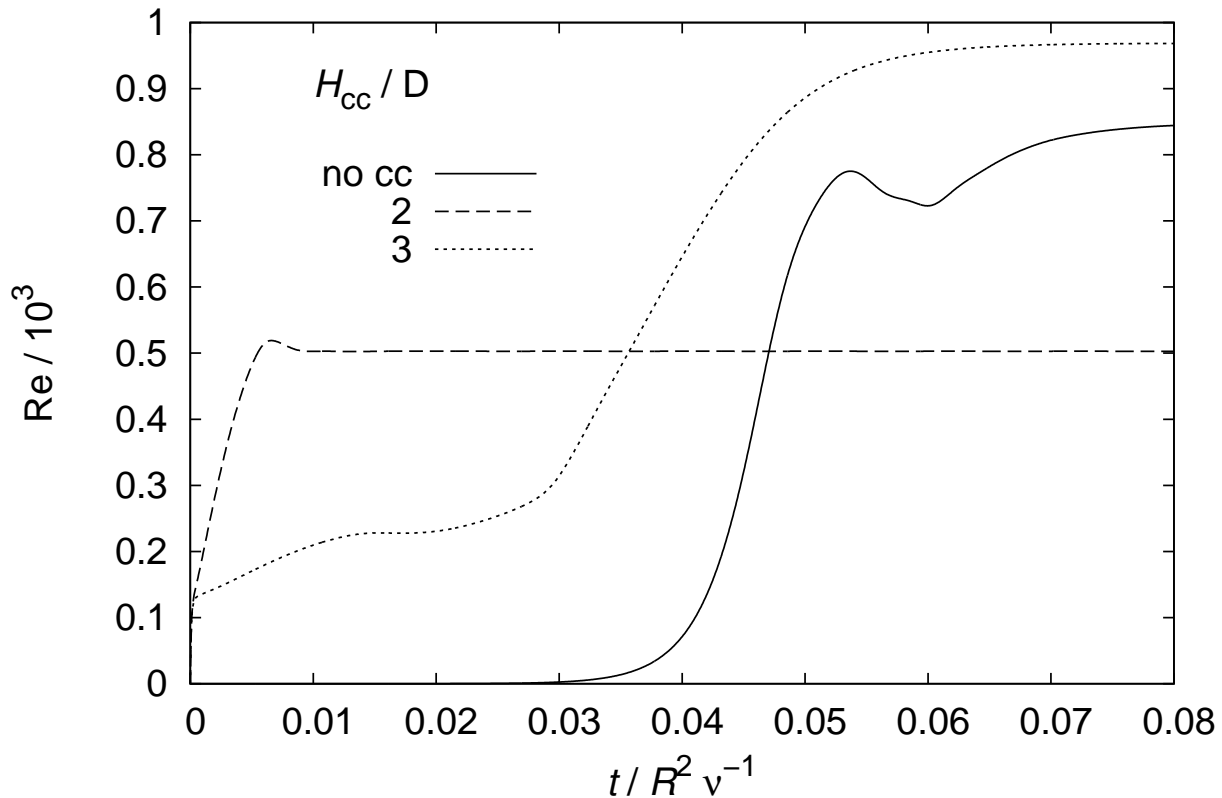}
\caption{\label{fig:fl:ti}Reynolds number in a cylindrical liquid metal column
($D=1$\,m, $H=2.4$\,m) for an applied current corresponding to
${\rm Ha=100}$. Simulation is carried out without current collector
(straight line) and a current collector of aspect ratio 2 (dashed
line) or 3 (dotted line) with an electric conductivity of $\sigma_{cc}/\sigma=5$.}
\end{figure}

Having clarified the geometry, grid resolution and initial fields,
we will investigate now the interaction of electro-vortex flows with
the TI. The TI usually growing exponentially needs some time to develop until
it reaches saturation (see the black line in Fig.~\ref{fig:fl:ti}).
We use this simulation of a liquid metal column of aspect ratio $H/D=2.4$
as reference and add current collectors of aspect ratio $H_{cc}/D=2$
(see the dashed line in Fig.~\ref{fig:fl:ti}). While the pure TI
evolves very slowly (black line), the electro-vortex flow (dashed
line) appears instantaneously at the start of the simulation. Nevertheless,
electro-vortex flow, which initially starts at the current collectors,
needs some time to spread over the whole volume. This spreading makes
the kinetic energy grow linearly until a certain saturation level
is attained. Although this energy is lower than the saturation level
of the TI, the latter does not develop here because the electro-vortex flow presumably disturbs the flow field
in such a way so that to suppress the TI . In order to study TI and
electro-vortex flow together, we increase the current collectors height
to an aspect ratio of $H_{cc}/D=3$ which lowers radial currents
and thus reduces the Lorentz force and the associated electro-vortex
flow (see the dotted line in Fig.~\ref{fig:fl:ti}). The first initial
rise of kinetic energy, which is the same as for short current collectors,
is followed by a much slower linear ``growth phase'' of the electro-
vortex flow. Of course, a weak flow needs more time to spread trough
the whole volume. After the electro-vortex flow reaches its saturation
level, the TI starts to develop and shortly an exponential growth
of energy can be observed. As expected, the final saturation level
lies slightly above the one of the simulation without current collectors
-- compare the straight and dotted line of Fig.~\ref{fig:fl:ti}.
Obviously, the TI arises only under favourable conditions: long and
good conducting current collectors, or possibly, also long liquid
metal columns.

\begin{figure}[t]
\centering{}
\includegraphics[width=\textwidth]{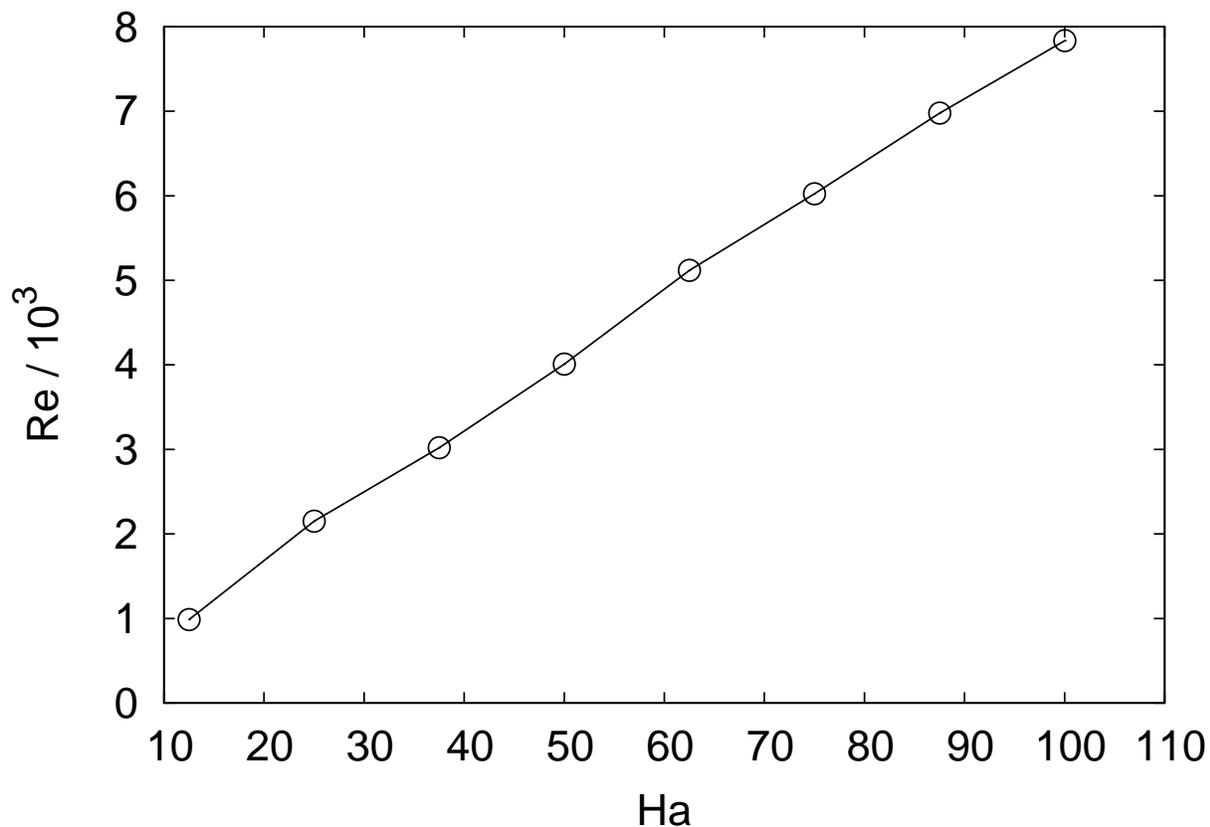}
\protect\caption{\label{fig:fl:i}Reynolds number of the electro-vortex flow against
the Hartmann number. The geometry consists of a cylindrical liquid
metal column ($D=1$\,m, $H=2.4$\,m) with two current collectors
($H_{cc}=0.5$\,m, $\sigma_{cc}/\sigma=5$).}
\end{figure}

In order to study the influence of ${\rm Ha}$, the aspect ratio and
conductivity of the current collectors on the electro-vortex flows we will
omit the TI in the following. Figure \ref{fig:fl:i} shows the Reynolds
number in dependence of the Hartmann number in a liquid metal column
of aspect ratio $H/D=2.4$ with relatively flat current collectors.
The relation is nearly linear. Compared to the TI, the magnitude of
electro-vortex flow is relatively high also for low Hartmann numbers.
In contrast to the TI, there is no critical current, i.e.\ already
arbitrary small Hartmann numbers will induce a fluid flow. Note also
that the relatively thick feeding line ($D_{fl}/D=0.5$) used here
and the high aspect ratio of the liquid metal cylinder ($H/D=2.4$)
may underestimate the electro-vortex flow taking place in a real liquid
metal battery.

\begin{figure}[t]
\centering{}
\includegraphics[width=1\columnwidth]{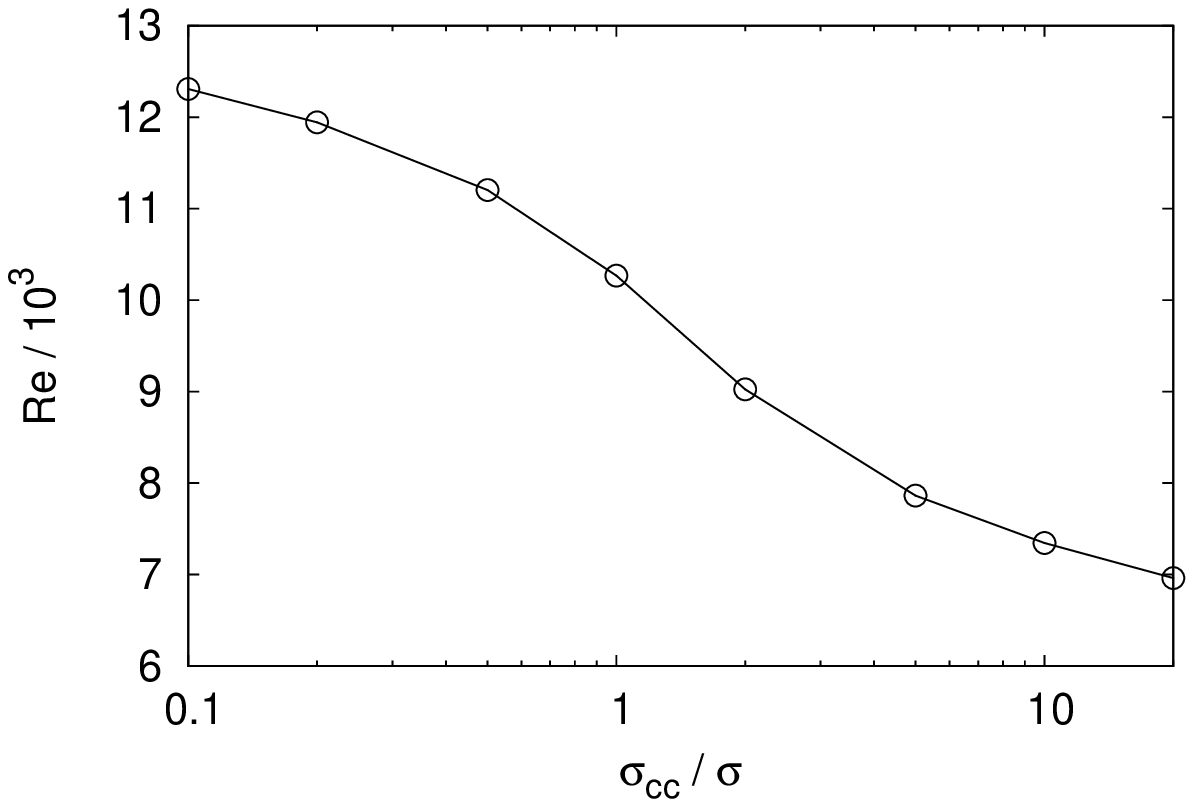}
\protect\caption{\label{fig:fl:sigma}Reynolds number of the electro-vortex flow (${\rm Ha}=100$)
in a cylindrical liquid metal column ($D=1$\,m, $H=2.4$\,m) with
two current collectors of aspect ratio $H_{cc}/D=0.5$. Only the electric
conductivity of the current collectors is changed.}
\end{figure}

With a constant current (${\rm Ha=100}$) and the geometry as before,
we alter the conductivity of the current collectors (see Fig.~\ref{fig:fl:sigma}).
Using perfectly conducting current collectors, the currents will homogenise
already in the current collector. Thus, we expect no radial currents
and very little Lorentz forces in the liquid metal. The curve in Fig.~\ref{fig:fl:sigma} will therefore approach asymptotically zero. Using
a poor conductor for the current collector leads to more radial currents
in the liquid. The increasing Lorentz forces will induce a larger
fluid flow. Concerning liquid metal batteries, Fig.~\ref{fig:fl:sigma}
reveals two things. Firstly, the current collector should have a good
conductivity in order to minimise material requirements. Secondly,
the conductivity has to be considered relative to that of the liquid
metal. The biggest changes are expected in the vicinity of
$\sigma_{cc}/\sigma=1$.

\begin{figure}[t]
\centering{}
\includegraphics[width=\textwidth]{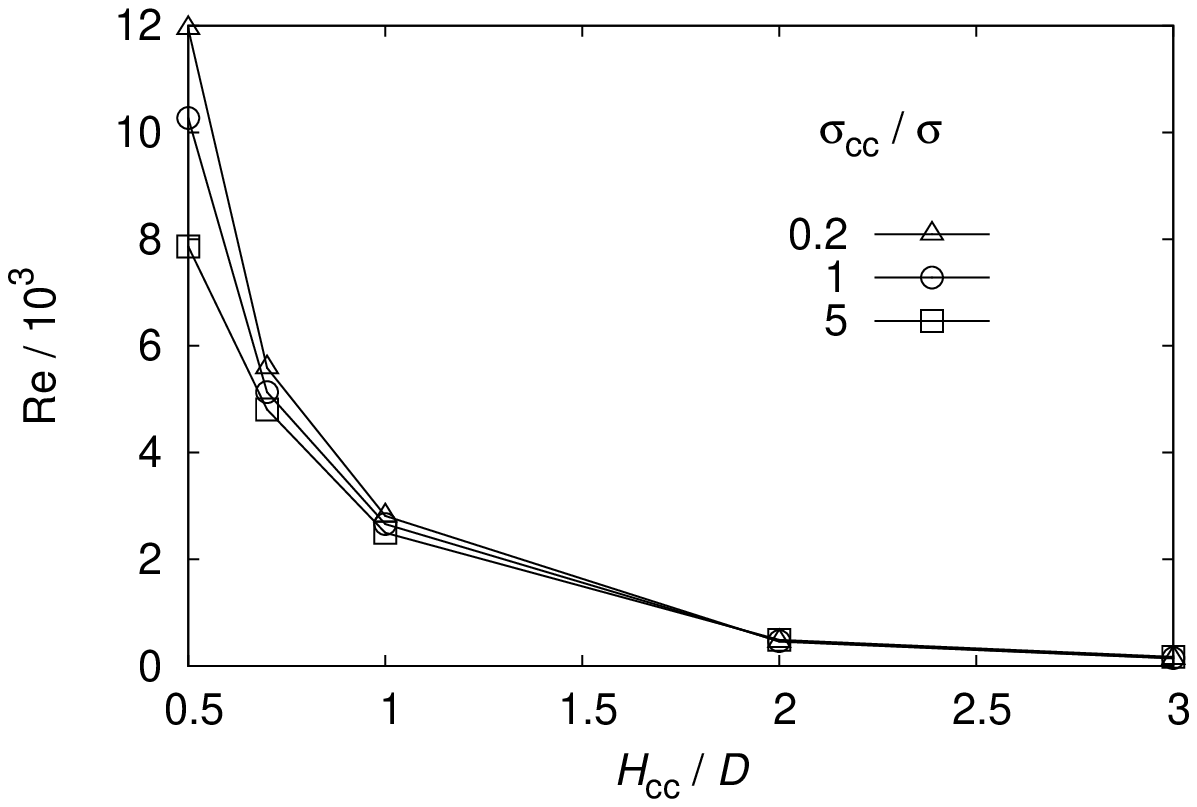}
\caption{\label{fig:fl:aspect}Reynolds number in a cylindrical liquid metal
column ($D=1$\,m, $H=2.4$\,m) for different current collector
aspect ratios and ${\rm Ha}=100$.}
\end{figure}

Assuming a fixed (dis-)charging current, the most important parameter determining
the flow velocity is the aspect ratio of current collectors (Fig.~\ref{fig:fl:aspect}). In the collectors of infinite height, the current
will redistribute uniformly producing no electro-vortex flow. With a
decrease of the aspect ratio, the kinetic energy of the liquid metal
increases exponentially. Especially going below an aspect ratio of
$H_{cc}/D=1$ produces a considerable flow in the fluid. The effect
of the aspect ratio of current collectors is important for flat collectors
but marginal for extended ones.

\section{Conclusions and prospects}

The first objective of this paper was to estimate the effects of the
current distribution in the solid current collectors on properties
of the Tayler instability in liquid metals. For this purpose, we have
enhanced the integro-differential equation approach, as developed
by Weber et al.\cite{Weber2013}, by a detailed computation of the electric potential
and the current distribution in the current collectors. This was accomplished
by solving the Laplace equation in the solid region and by matching
this solution in an iterative way to the solution of the Poisson equation
in the liquid metal.

Our main conclusion is that the consideration of realistic current
collectors has certain effects on the TI which make them essential
for a realistic simulation. However, in a real fluid flow, thermal
convection and electro-vortex flow will have a much larger influence
on the TI. Therefore, it seems somehow justified to work with simplified
boundary conditions such as $\varphi=\textit{const}$ for a first
estimate, bearing in mind the underestimation of the growth rate and
flow velocity.

Secondly, we have included the feeding lines into our simulation by
solving a coupled Laplace equation for the static potential in the
liquid metal and current collectors. Flowing from the thin wire into
the wide current collector, the electric current acquires radial components,
resulting in electro-vortex flow. Depending on the geometry of battery,
current collectors and their conductivities, these flows can be quite
strong and may impede the development of the TI.

In LMBs, strong electro-vortex flows have to be avoided in order to
maintain the integrity of the thin electrolyte layer. This may by
achieved by an appropriate design of the current collectors. Of course,
a solid metal cylinder with an aspect ratio $H/D=1$ will get heavy
and costly, but a careful distribution of the charging current by
several (thin) wires may solve the problem. A more detailed design
will require further studies under the cost aspect.

While a strong mixing of the LMB must be avoided, a slight fluid flow
in the lower compartment can be advantageous. Reaction products, as,
e.g.\ intermetallic phases which would limit the mass transfer at the
interface may be spread in the alloy by a slow fluid flow. Again,
this can be achieved by an appropriate geometry of the positive current
collector. Here, not only aspect ratio and conductivity will be of
interest, but also surface shaping, coating, etc.\ are possible options.
Further numerical and experimental studies will be required.

Apart from studying the TI and electro-vortex flow, the methodology
developed here can be useful when it comes to a realistic simulation
of the interaction of the TI in the bulk of the fluid with various
types of interfacial instabilities between the layers of a LMB. Work
on this problem is currently under way.

\section*{Acknowledgements}
This work was supported by Helmholtz-Gemeinschaft Deutscher
Forschungszentren (HGF) in frame of the ``Initiative für mobile und
stationäre Energiespeichersysteme'', and in frame of the Helmholtz
Alliance LIMTECH, as well as by Deutsche Forschungsgemeinschaft in
frame of the SPP 1488 (PlanetMag).

\end{document}